\documentclass{aa}  

\usepackage{graphicx}
\usepackage{txfonts}
\defcitealias{Dudi10}{D10}
\begin{document}

   \title{Liverpool-Maidanak monitoring of the Einstein Cross in 2006$-$2019}

   \subtitle{I. Light curves in the $gVrRI$ optical bands and microlensing signatures\thanks{Tables 
    4$-$8 and 10$-$14 are only available in electronic form at the CDS via anonymous ftp to 
    cdsarc.u-strasbg.fr (130.79.128.5) or via http://cdsweb.u-strasbg.fr/cgi-bin/qcat?J/A+A/vol/page}}    
                                                             
   \author{L. J. Goicoechea\inst{1}
                \and
                B. P. Artamonov\inst{2}
                \and 
                V. N. Shalyapin\inst{1,3,4}
                \and
                A. V. Sergeyev\inst{4,5}
                \and
                O. A. Burkhonov\inst{6}
                \and
                T. A. Akhunov\inst{7}
                \and
                I. M. Asfandiyarov\inst{6}
                \and
                V. V. Bruevich\inst{2}
                \and
                S. A. Ehgamberdiev\inst{6}
                \and
                E. V. Shimanovskaya\inst{2}
                \and
                A. P. Zheleznyak\inst{4}        
                }

\institute{Departamento de F\'\i sica Moderna, Universidad de Cantabria, 
                Avda. de Los Castros s/n, E-39005 Santander, Spain\\
                \email{goicol@unican.es;vshal@ukr.net}
                \and
                Sternberg Astronomical Institute, Lomonosov Moscow State University,
                Universitetsky pr. 13, 119992 Moscow, Russia\\
                \email{bartamon@mail.ru}
                \and
                O.Ya. Usikov Institute for Radiophysics and Electronics, National 
                Academy of Sciences of Ukraine, 12 Acad. Proscury St., UA-61085 
                Kharkiv, Ukraine
                \and
                Institute of Astronomy of V.N. Karazin Kharkiv National University,
                Svobody Sq. 4, UA-61022 Kharkiv, Ukraine
                \and
		    Institute of Radio Astronomy of the National Academy of Sciences of 
                Ukraine, 4 Mystetstv St., UA-61002 Kharkiv, Ukraine
		    \and
                Ulugh Beg Astronomical Institute of the Uzbek Academy of Sciences,
                Astronomicheskaya 33, 100052 Tashkent, Uzbekistan
                \and
                National University of Uzbekistan, Department of Astronomy and 
                Atmospheric Physics, 100174 Tashkent, Uzbekistan}

%   \date{Received September 15, 1996; accepted March 16, 1997}

% \abstract{}{}{}{}{} 
% 5 {} token are mandatory
 
  \abstract{Quasar microlensing offers a unique opportunity to resolve tiny sources in 
distant active galactic nuclei and study compact object populations in lensing 
galaxies. We therefore 
searched for microlensing-induced variability of the gravitationally lensed quasar QSO 
2237+0305 (Einstein Cross) using 4\,374 optical frames taken with the 2.0 m Liverpool 
Telescope and the 1.5 m Maidanak Telescope. These $gVrRI$ frames over the 2006$-$2019 
period were homogeneously processed to generate accurate long-term multi-band light 
curves of the four quasar images A-D. Through difference light curves, we found strong 
microlensing signatures. We then focused on the analytical modelling of two putative 
caustic-crossing events in image C, finding compelling evidence that this image 
experienced a double caustic crossing. Additionally, our overall results 
indicate that a standard accretion disc accounts reasonably well for the brightness profile 
of UV continuum emission sources and for the growth in source radius when the emission 
wavelength increases: $R_{\lambda} \propto \lambda^{\alpha}$, $\alpha$ = 1.33 $\pm$ 
0.09. However, we caution that numerical microlensing simulations are required before 
firm conclusions can be reached on the UV emission scenario because the $VRI$-band 
monitoring during the first caustic crossing and one of our two $\alpha$ indicators 
lead to a few good solutions with $\alpha \approx$ 1.}     
  

   \maketitle
%
%-------------------------------------------------------------------

\section{Introduction}
\label{sec:intro}

Analysis of multiply imaged quasars (which undergo strong gravitational lensing) 
reveals the structure and composition of tiny regions in distant active galactic 
nuclei, halos of intervening galaxies, and intergalactic space 
\citep[e.g.][]{Schn92,Schn06}. For a given gravitationally lensed quasar, multi-band
optical light curves of its multiple images sometimes show phases of chromatic 
microlensing activity. This activity is related to microlenses (stars) that affect each 
continuum-emitting region differently and to a different extent, so that more compact 
(bluer) sources are expected to suffer stronger effects \citep[e.g.][and references 
therein]{Mosq11}. Therefore multi-band photometric monitorings of microlensing episodes in 
lensed quasars are used, among other things, to probe the relationship between 
source radius and emission wavelength $\lambda$. Although some multi-band light curves 
only provided evidence that bluer sources are smaller \citep[e.g.][]{Vaku04}, 
several studies indicated that microlensing-induced chromatic variations are fully or 
marginally consistent with radii that grow as $\lambda^{\alpha}$, $\alpha$ = 4/3 
\citep[standard disc model; e.g.][]{Shal02,Angu08,Eige08,Poin08,Hain13,Blac15,Mun16}. 

\object{QSO 2237+0305} (the Einstein Cross; $z_{\rm{s}}$ = 1.695) consists of four 
quasar images (A, B, C, and D) that are arranged like a cross around the nucleus of a 
nearly face-on spiral galaxy at $z_{\rm{l}}$ = 0.039 \citep{Huch85,Yee88}. Although 
the light of this quadruply imaged quasar passes through four different regions in the 
bulge of the lensing spiral galaxy, time delays between images are extraordinarily 
short \citep[typical values range from a few hours to a few days; 
e.g.][]{Schn88,Vaku06}. As a result of these short delays, magnitude differences 
between any two images exclusively include microlensing variations because intrinsic 
variations are removed \citep{Irwi89}. Taking advantage of this fact, \citet{Eige08} used 
the chromaticity of $A - B$ to robustly constrain the power-law index $\alpha$. They 
analysed a three-year spectroscopic monitoring at the European Southern Observatory (ESO) by 
removing the broad emission lines and the iron pseudo-continuum from quasar spectra in 
39 epochs and by focusing on the continuum for $\lambda$ values in the 1500$-$3000 \AA\ 
interval. Each spectral distribution of the continuum was then integrated in six independent
250 \AA\ bands to construct multi-band light curves. The six brightness 
records for A and B, along with the well-sampled $V$-band light curves of both images 
from the Optical Gravitational Lensing Experiment \citep[OGLE;][]{Wozn00,Udal06}, led 
to $\alpha$ = 1.2 $\pm$ 0.3, in good agreement with a standard accretion disc. This 
result agrees with that of \citet{Mun16} from a follow-up in six narrow bands in six 
epochs ($\alpha \approx$ 1.0 $\pm$ 0.3), which is only marginally consistent 
with a standard disc, however.  

In addition to the growth of the source radius with increasing emission wavelength, 
the surface 
brightness profile of sources at different wavelengths is a key piece to understand 
the accretion disc structure. While it has been proved that the shape of this profile 
does not play a relevant role in accounting for microlensing effects away from 
(micro)caustics \citep[e.g.][]{Mort05}, well-sampled light curves of \object{QSO 
2237+0305} are sensitive to the size and shape of emission regions when these 
regions cross caustic folds \citep[e.g.][]{Shal02,Gilm06,Kopt07a,Abol12,Medi15}, 
favouring the 
standard disc profile or its relativistic version. Except for the Gravitational Lenses 
International Time Project (GLITP) observations in the $R$ band \citep{Alca02}, the 
only finely sampled light curves of the Einstein Cross that have been deeply 
interpreted are those of the GLITP and OGLE collaborations in the $V$ band (source 
emitting at $\lambda \sim$ 2000 \AA). These $V$-band records provided not only 
constraints on the source geometry, but also information on physical properties of the 
lensing galaxy \citep[fraction of mass in stars, mean stellar mass, and transverse 
velocity; e.g.][]{Koch04,Gilm05}. Even former poorly sampled light curves 
\citep{Irwi89,Corr91,Oste96} led to interesting physical constraints 
\citep[e.g.][]{Wyit99,Wyit00a,Wyit00b}. Therefore, new well-sampled multi-band light 
curves of \object{QSO 2237+0305} are promising tools for improving our knowledge of the 
distant active galactic nucleus and the local intervening spiral galaxy.  

This paper describes a collaborative project that analysed optical frames of the Einstein 
Cross in a homogeneous way (using the same photometric method), built accurate 
multi-band light curves of the four quasar images throughout the last 14 years 
(2006$-$2019), and searched for new microlensing-induced variations. The project relied 
on a large set of $gVrRI$ frames taken from two telescopes in the northern hemisphere: 
the 2.0 m Liverpool Telescope (LT; using $gr$ Sloan filters) and the 1.5 m 
telescope at the Maidanak Observatory (hereafter MT; using $VRI$ Bessell filters). In 
Sect.~\ref{sec:obslcur} we present the 14-year multi-band monitoring with the LT and 
the MT, outline main photometric tasks required to extract quasar fluxes, and show new and 
updated light curves of A, B, C, and D. In Sect.~\ref{sec:diflcur} we discuss 
microlensing signatures in difference light curves and focus on a possible double
caustic-crossing event (DCCE) in C. Our conclusions are summarised in 
Sect.~\ref{sec:end}. 

\section{Observations and data reduction}
\label{sec:obslcur}

First $VRI$ photometric observations of \object{QSO 2237+0305} with the MT were 
performed in 1995 \citep{Vaku97}, and $VRI$ light curves over the first monitoring 
decade have been described in several previous papers \citep[e.g.][]{Vaku04,Kopt07b}. Here,
we present new MT observations from 2006 to 2019\footnote{Observations in 2006$-$2008 
have previously been presented in an Ukrainian journal \citep{Dudi10}}. An important 
upgrade of the 
telescope occurred in 2006 by installing the SNUCAM camera, which uses a CCD detector 
with a $0\farcs266$ pixel$^{-1}$ scale \citep{Im10}. Although this camera is still 
working on the MT, we used the FLI MicroLine CCD with a pixel scale 
of $0\farcs21$ in 2012 and 2017. During the new observing period, we collected frames 
in $VRI$ Bessell 
passbands. This translates into a follow-up of sources emitting at effective wavelengths 
$\lambda_V$ = 2002 \AA, $\lambda_R$ = 2398 \AA, and $\lambda_I$ = 3113 \AA. Before 
 quasar fluxes were extracted, basic instrumental reductions were applied to all MT frames. 
This incorporated bias subtraction, dark frame subtraction (only for FLI MicroLine 
data), trimming of the overscan regions, flat fielding, and cosmic-ray cleaning. 
Moreover, we mapped pixel instrumental locations to their positions in the World 
Coordinate System, inserting sky coordinates into frame headers.   

Additionally, the monitoring with the LT in $gr$ Sloan bands started in 2006, soon 
after the commencement of science operations for this robotic telescope \citep{Ste04}, 
and \citet{Gilm18} have shown $r$-band light curves over two four-year periods. In this 
paper, we describe the full database between 2006 and 2019, including new LT 
observations in the $g$ band, as well as an extended (updated) set of frames in the 
$r$ band. Frames in 2006$-$2009 were taken with the RATCam CCD camera ($0\farcs27$ 
pixel$^{-1}$ scale), whereas we used the IO:O CCD camera ($0\farcs30$ pixel$^{-1}$ 
scale) from 2013 onwards. Regarding effective wavelengths in the quasar rest frame, we 
have $\lambda_g$ = 1779 \AA\ and $\lambda_r$ = 2296 \AA. In addition to basic 
pre-processing tasks included in the LT pipelines, we cleaned cosmic rays and 
interpolated over bad pixels using bad-pixel masks. Many LT frames of \object{QSO 
2237+0305} were already incorporated into the Gravitational LENses and DArk MAtter 
(GLENDAMA) database\footnote{\url{https://grupos.unican.es/glendama/database}} 
\citep{Gilm18}, and the next update of this archive will allow us to add all available 
LT-MT data of the Einstein Cross. The summary of LT-MT observations is shown in 
Table~\ref{tab:obs}. 

Point-spread function (PSF) fitting photometry is particularly useful to extract 
fluxes of closely spaced quasar images. This photometric method relies on the 
assumption that all point-like sources can be represented by the same PSF, which is 
well traced by an analytical function or a field star \citep[e.g.][]{Howe06}. We 
performed PSF-fitting photometry on the Einstein Cross using the 2D profile of a 
field star as empirical PSF (see details on field stars in Table~\ref{tab:stars}). The 
brightest star ($\gamma$) was used as PSF in most frames. However, when $\gamma$ was 
saturated or had defective pixels, we took the PSF of the star $\alpha$. The $\alpha$ 
star also served for estimating the signal-to-noise ratio ($S/N$) in each frame, and to 
calculate magnitude zero-points. PSF-fitting photometry on the $\beta$ star was 
used to verify that the quasar variability is real.

\begin{table*}
\centering
\caption{Liverpool-Maidanak monitoring of QSO 2237+0305 in 2006$-$2019.}
\begin{tabular}{ccccccc}
\hline\hline
Band & Total frames & Selected frames\tablefootmark{a} & 
Epochs/nights & $\langle T_{\rm{frame}} \rangle$\tablefootmark{b} (s) &
$\langle T_{\rm{night}} \rangle$\tablefootmark{c} (s) & 
$FWHM$\tablefootmark{d} (\arcsec) \\
\hline
$g$ &  277 &  260 & 203 & 253 &  318 & 1.54 $\pm$ 0.16 \\     
$V$ &  731 &  695 & 180 & 290 & 1124 & 1.26 $\pm$ 0.14 \\
$r$ &  366 &  328 & 253 & 200 &  255 & 1.46 $\pm$ 0.15 \\
$R$ & 2295 & 2192 & 445 & 260 & 1290 & 1.25 $\pm$ 0.15 \\
$I$ &  705 &  678 & 179 & 200 &  758 & 1.17 $\pm$ 0.13 \\
\hline
\end{tabular}
\tablefoot{
\tablefoottext{a}{Number of individual frames after removing those with relatively 
poor quality. The removed frames produce anomalous photometric results and account for 
4$-$10\% of the total in each optical band};
\tablefoottext{b}{average exposure time per individual frame};
\tablefoottext{c}{average exposure time per night};
\tablefoottext{d}{mean value and standard deviation of the full width at half-maximum 
($FWHM$) of the seeing disc.} 
} 
\label{tab:obs}
\end{table*}

\begin{table*}
\centering
\caption{Stars around QSO 2237+0305.}
\begin{tabular}{cccccccc}
\hline\hline
Star & RA(J2000) & Dec(J2000) & $g$ & $V$ & $r$ & $R$ & $I$ \\
\hline
$\alpha$ & 340.10997 & 3.35061 & 17.795 & 17.500 & 17.209 & 17.280 & 17.260 \\
$\beta$  & 340.11231 & 3.37664 & 18.684 & 18.276 & 17.805 & 17.855 & 17.692 \\
$\gamma$ & 340.12079 & 3.33275 & 16.074 & 15.813 & 15.583 & 15.661 & 15.690 \\
\hline
\end{tabular}
\tablefoot{
Field stars named $\alpha$ and $\beta$ are shown in Fig. 1 of \citet{Corr91}, and the
$\gamma$ star is the so-called star 1 in \citet{More05}. RA(J2000) and Dec(J2000) are 
given in degrees. The $gr$ magnitudes are taken from the Sloan Digital Sky Survey DR15 
\citep{Agua19}, while the $VRI$ magnitudes correspond to results in 
\citet{Corr91} and \citet{Vaku97}. 
} 
\label{tab:stars}
\end{table*}

\begin{table*}
\centering
\caption{Structural parameters of the lensing galaxy.}
\begin{tabular}{ccccc}
\hline\hline
Band & Relative flux & Effective radius (\arcsec) & $b/a$ & $PA$ (\degr) \\
\hline
$g$ & 1.29 $\pm$ 0.11 & 5.71 $\pm$ 0.39 & 0.627 $\pm$ 0.024 & 64.6 $\pm$ 1.2 \\
$V$ & 1.36 $\pm$ 0.22 & 5.11 $\pm$ 0.58 & 0.643 $\pm$ 0.034 & 66.5 $\pm$ 1.2 \\
$r$ & 1.68 $\pm$ 0.13 & 5.15 $\pm$ 0.36 & 0.619 $\pm$ 0.021 & 64.4 $\pm$ 0.9 \\
$R$ & 1.53 $\pm$ 0.06 & 4.52 $\pm$ 0.16 & 0.616 $\pm$ 0.011 & 65.7 $\pm$ 0.6 \\
$I$ & 2.04 $\pm$ 0.15 & 4.29 $\pm$ 0.26 & 0.629 $\pm$ 0.018 & 66.4 $\pm$ 0.9 \\
\hline
\end{tabular}
\tablefoot{
Mean value and standard deviation of the relative flux (galaxy-to-$\gamma$ stellar flux 
ratio), effective radius, axis ratio ($b/a$), and orientation ($PA$) for a de 
Vaucouleurs profile. 
} 
\label{tab:gal}
\end{table*}

In the crowded region containing the four images of \object{QSO 2237+0305} (QSO 
subframes), the photometric model consisted of a constant background, four point-like 
sources, and a de Vaucouleurs profile convolved with the PSF \citep{Alca02,Gilm18}. 
This last ingredient accounts for the light distribution of the lensing galaxy bulge. 
Taking the position of image A as a reference for astrometry, and setting the relative 
positions of B-D and the centre of the galaxy to those obtained from {\it 
Hubble} Space Telescope (HST) data in the $H$ band \citep[e.g. Table 1 of][]{Alca02}, 
we fitted the model to each QSO subframe using the IMFITFITS software\footnote{The 
IMFITFITS code minimises the sum of squared residuals.} \citep{McLe98}. Our initial 
model had 11 free parameters: 2D position of A, sky background, galaxy structural 
parameters (flux, effective radius, axis ratio, and orientation) and four quasar 
fluxes. It was only applied to a large set of good frames in terms of seeing and $S/N$. 
Results from this initial iteration allowed us to determine the inner structure of the 
galaxy (see Table~\ref{tab:gal}). In a second iteration, we applied IMFITFITS to all 
frames, setting relative positions and galaxy parameters. A number of individual 
frames produced anomalous photometric results (outliers). These are characterised by a 
poor image quality and were therefore removed from the final database. Additionally, 
we used the 
simplest photometric model (point-like source plus constant background) to extract 
fluxes of the $\beta$ control star. 

\setcounter{table}{8}
\begin{table}
\centering
\caption{Mean magnitude errors of the quasar and control star.}
\begin{tabular}{ccccccc}
\hline\hline
Band & N$_{\rm{pairs}}$\tablefootmark{a} & A & B & C & D & $\beta$ \\
\hline
$g$ &  45 & 0.013 & 0.019 & 0.053 & 0.022 & 0.013 \\
$V$ &  57 & 0.014 & 0.013 & 0.029 & 0.022 & 0.017 \\
$r$ &  76 & 0.013 & 0.028 & 0.049 & 0.029 & 0.010 \\
$R$ & 229 & 0.015 & 0.017 & 0.033 & 0.024 & 0.015 \\
$I$ &  58 & 0.012 & 0.016 & 0.026 & 0.017 & 0.014 \\
\hline
\end{tabular}
\tablefoot{
\tablefoottext{a}{Number of pairs of magnitudes that we used to estimate the mean 
uncertainties.}
} 
\label{tab:error}
\end{table}

Detailed photometric results for all individual frames are available in Tables 4 ($g$ 
band), 5 ($V$ band), 6 ($r$ band), 7 ($R$ band), and 8 ($I$ band) at the CDS: Cols. 
1$-$12 list the civil date and frame number on that date (yymmdd\_number), the 
observing epoch (MJD$-$50\,000), the exposure time (s), $FWHM$ (\arcsec), the PSF 
ellipticity, $S/N$, $A$ (mag), $B$ (mag), $C$ (mag), $D$ (mag), $\beta$ (mag), and the 
reduced chi-square ($\chi^2$/dof, where 'dof' denotes the degrees of freedom) value 
when IMFITFITS was applied on the QSO subframe, respectively. Column 13 contains an 
asterisk for removed 
poor-quality frames or is empty for selected (non-removed) frames. Results for 
selected frames were combined on a nightly basis to obtain magnitudes at 203 ($g$ 
band), 180 ($V$ band), 253 ($r$ band), 445 ($R$ band), and 179 ($I$ band) epochs (see 
Table~\ref{tab:obs}). To estimate mean photometric errors in the light curves of A-D 
and $\beta$, we calculated deviations between adjacent magnitudes that are separated 
from each other by no more than 2.5 d. For each optical band, the number of pairs used 
and the mean deviations are displayed in Table~\ref{tab:error}. The typical 
uncertainties in quasar magnitudes range from $\sim$1\% for the brightest (A) image to 
$\sim$2$-$5\% for the generally faintest (C) image. For a given band, errors at every 
epoch were then computed by weighting mean values by the $\langle S/N \rangle / S/N$ 
ratio \citep[e.g.][]{Howe06}. 

\begin{figure*}
\centering
\includegraphics[width=9cm]{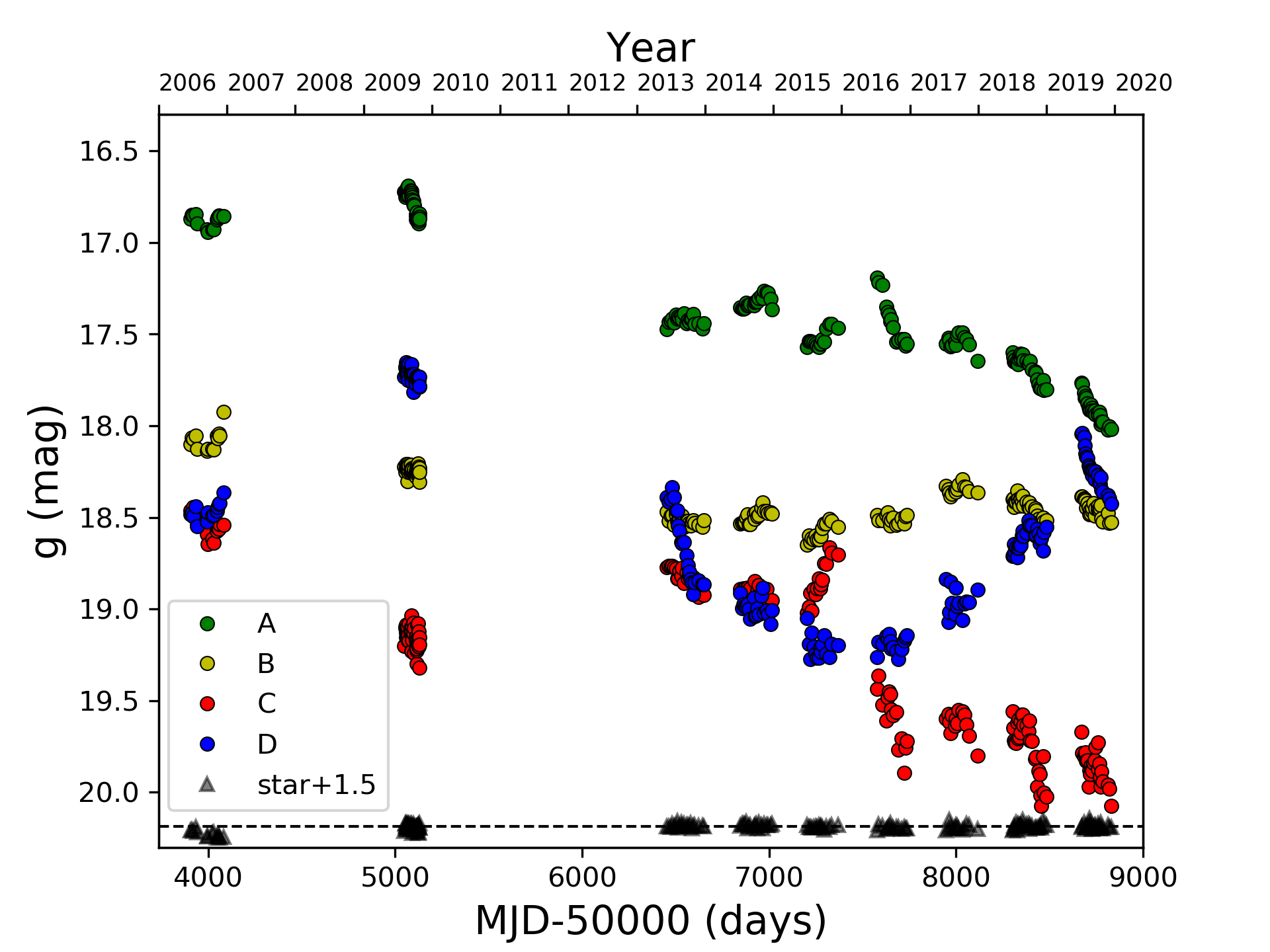}
\includegraphics[width=9cm]{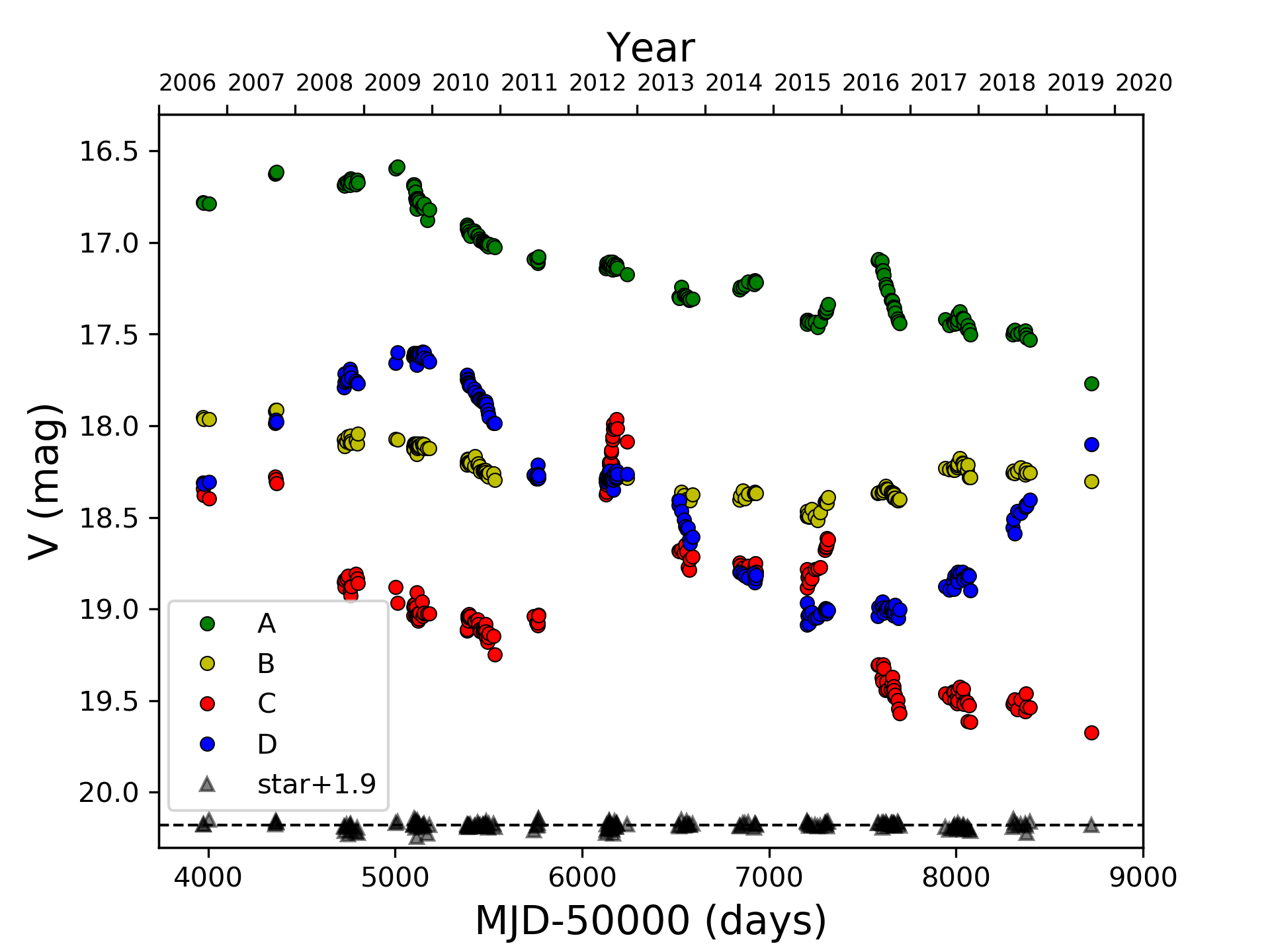}
\includegraphics[width=9cm]{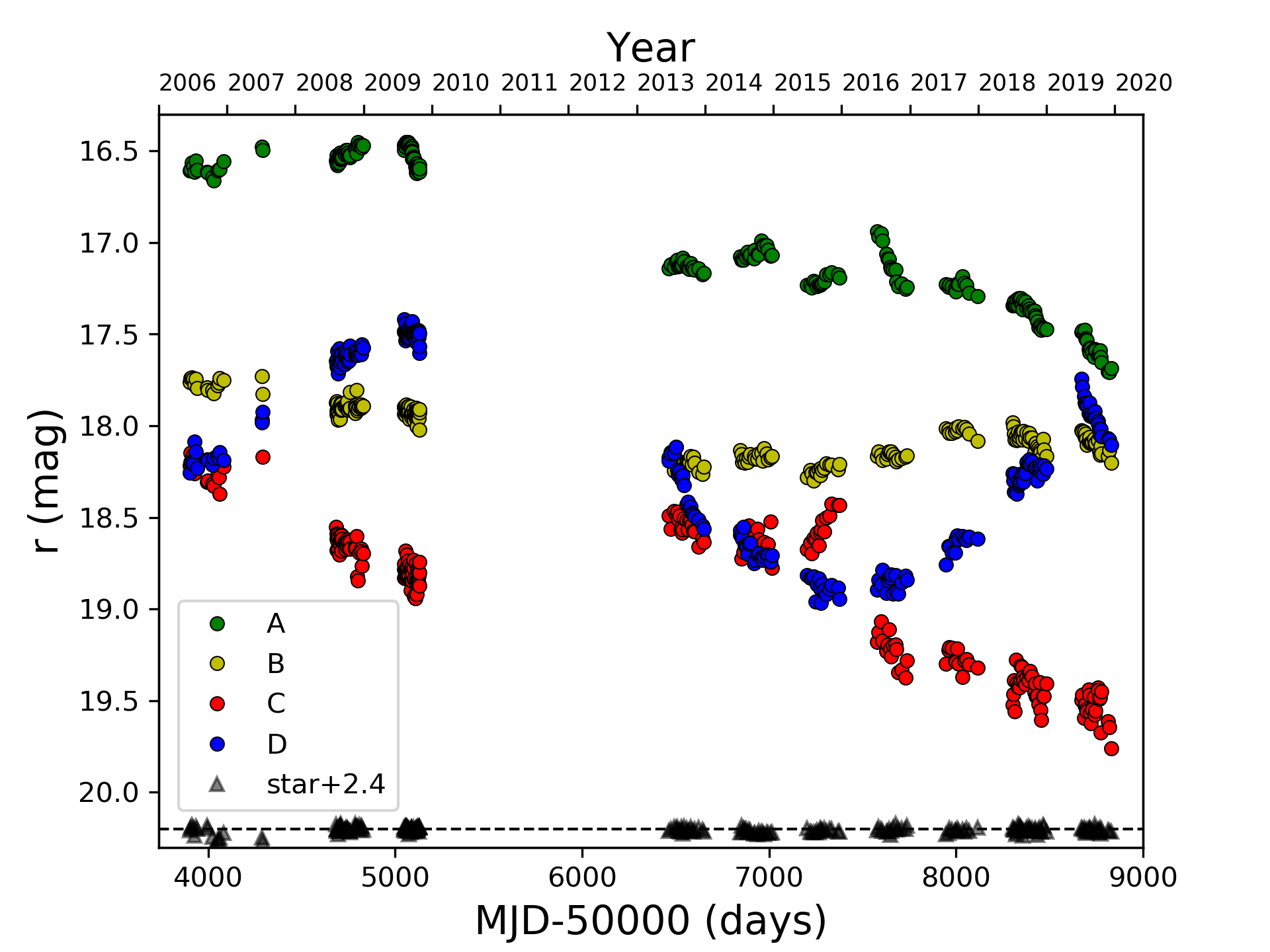}
\includegraphics[width=9cm]{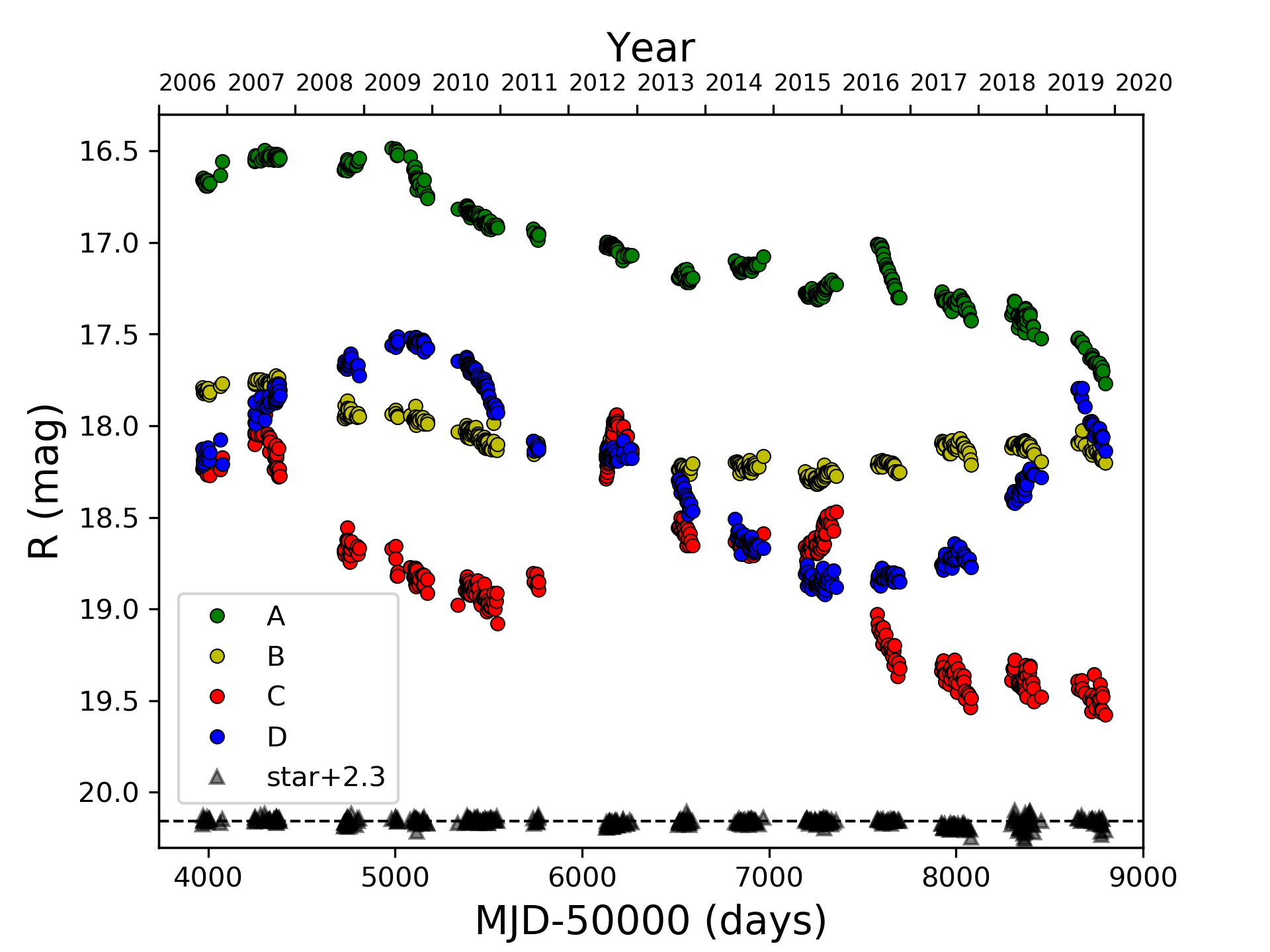}
\includegraphics[width=9cm]{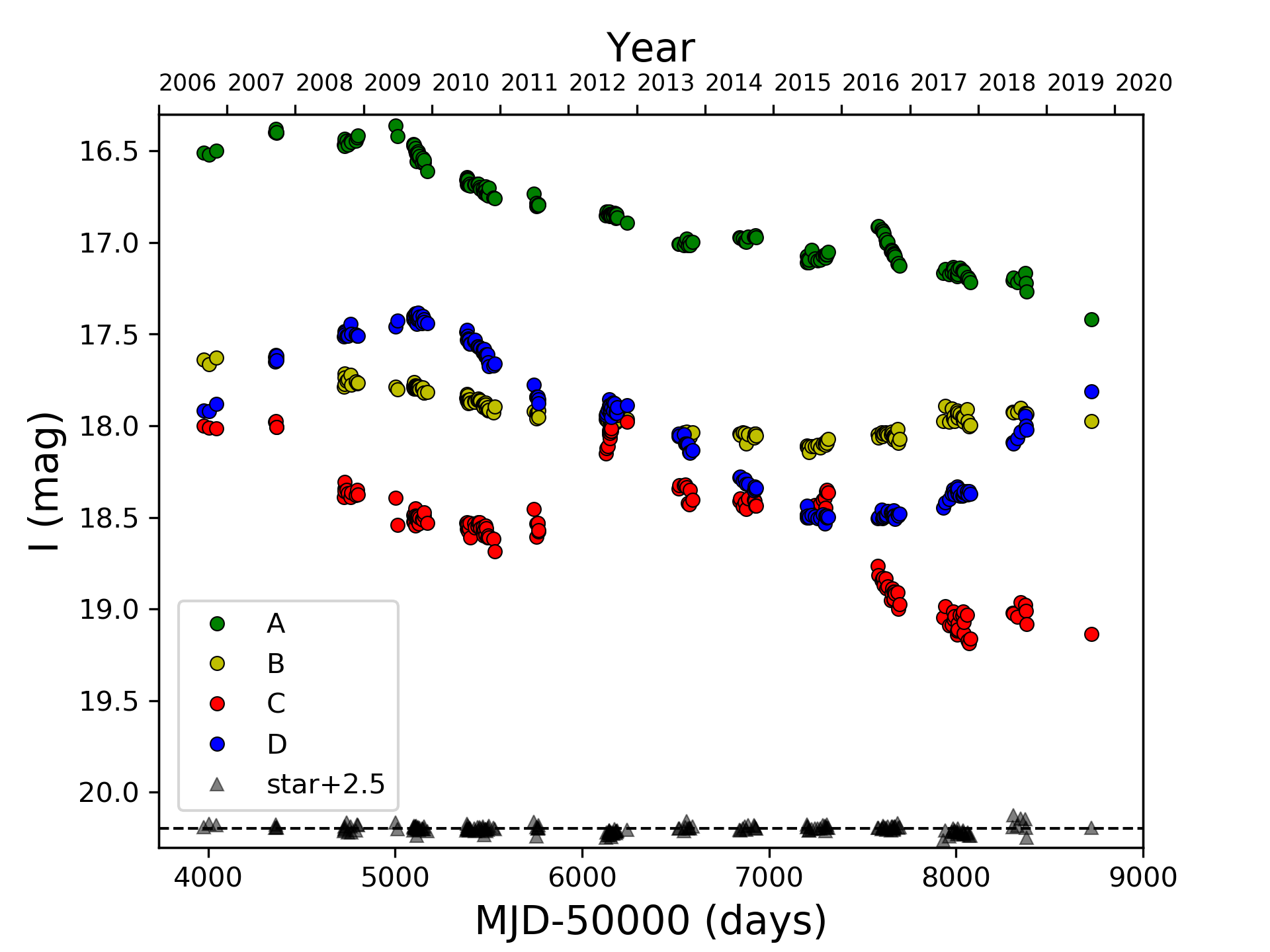}
\caption{Liverpool-Maidanak light curves of \object{QSO 2237+0305}. The quasar 
variability is measured at effective rest-frame wavelengths in the UV region 
($\sim$1780$-$3110 \AA) over a 14-year period.}
\label{fig:lcur}
\end{figure*}

The final light curves of A-D and $\beta$ are shown in Figure~\ref{fig:lcur}, and they 
are provided in tabular format at the CDS\footnote{See also 
\url{https://grupos.unican.es/glendama/q2237.htm} for updated results}: Tables 10 ($g$ 
band), 11 ($V$ band), 12 ($r$ band), 13 ($R$ band), and 14 ($I$ band). These five 
machine-readable ASCII files are structured in the same manner. Column 1 includes 
the observing epoch (MJD$-$50\,000), while Cols. 2$-$3, 4$-$5, 6$-$7, 8$-$9, and 
10$-$11 display the magnitudes and magnitude errors of A, B, C, D, and $\beta$, 
respectively. In Appendix~\ref{sec:appena} we compare our $V$-band magnitudes in 
2006$-$2008 with those obtained through a different photometric technique
and the OGLE data at the same epochs. 
    
\section{Difference light curves: new tools for microlensing studies}
\label{sec:diflcur}

\begin{figure*}
\centering
\includegraphics[width=9cm]{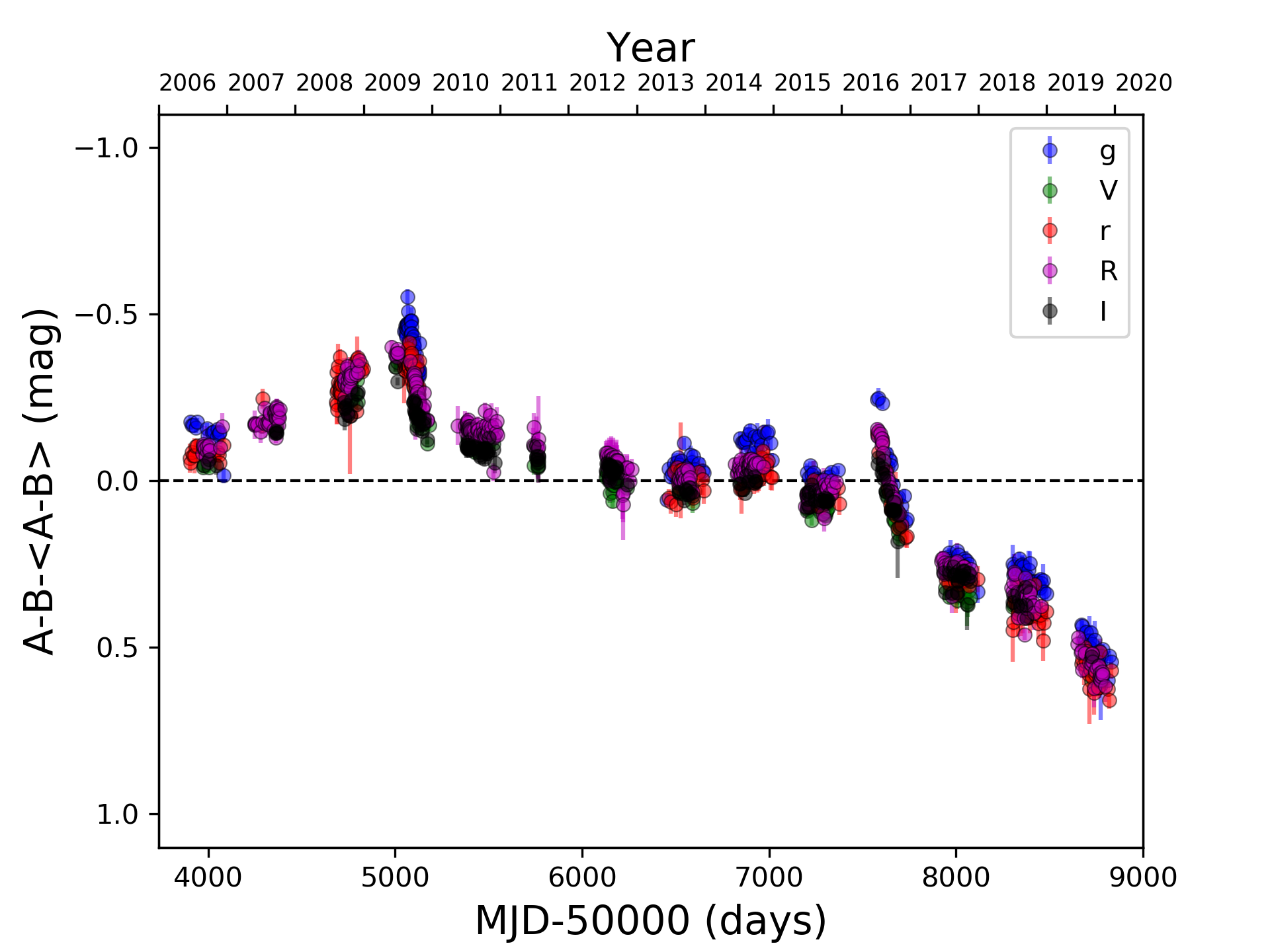}
\includegraphics[width=9cm]{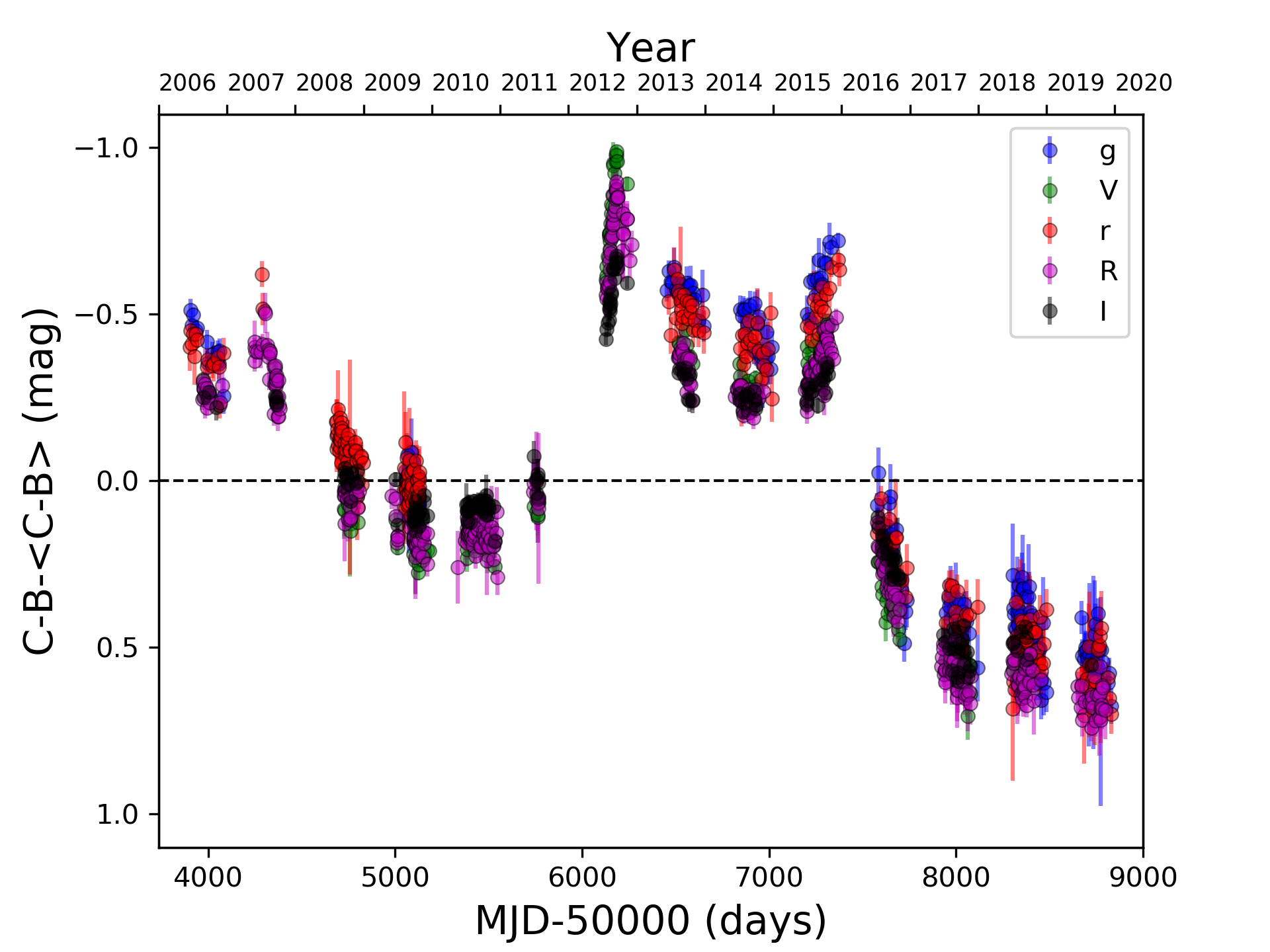}
\includegraphics[width=9cm]{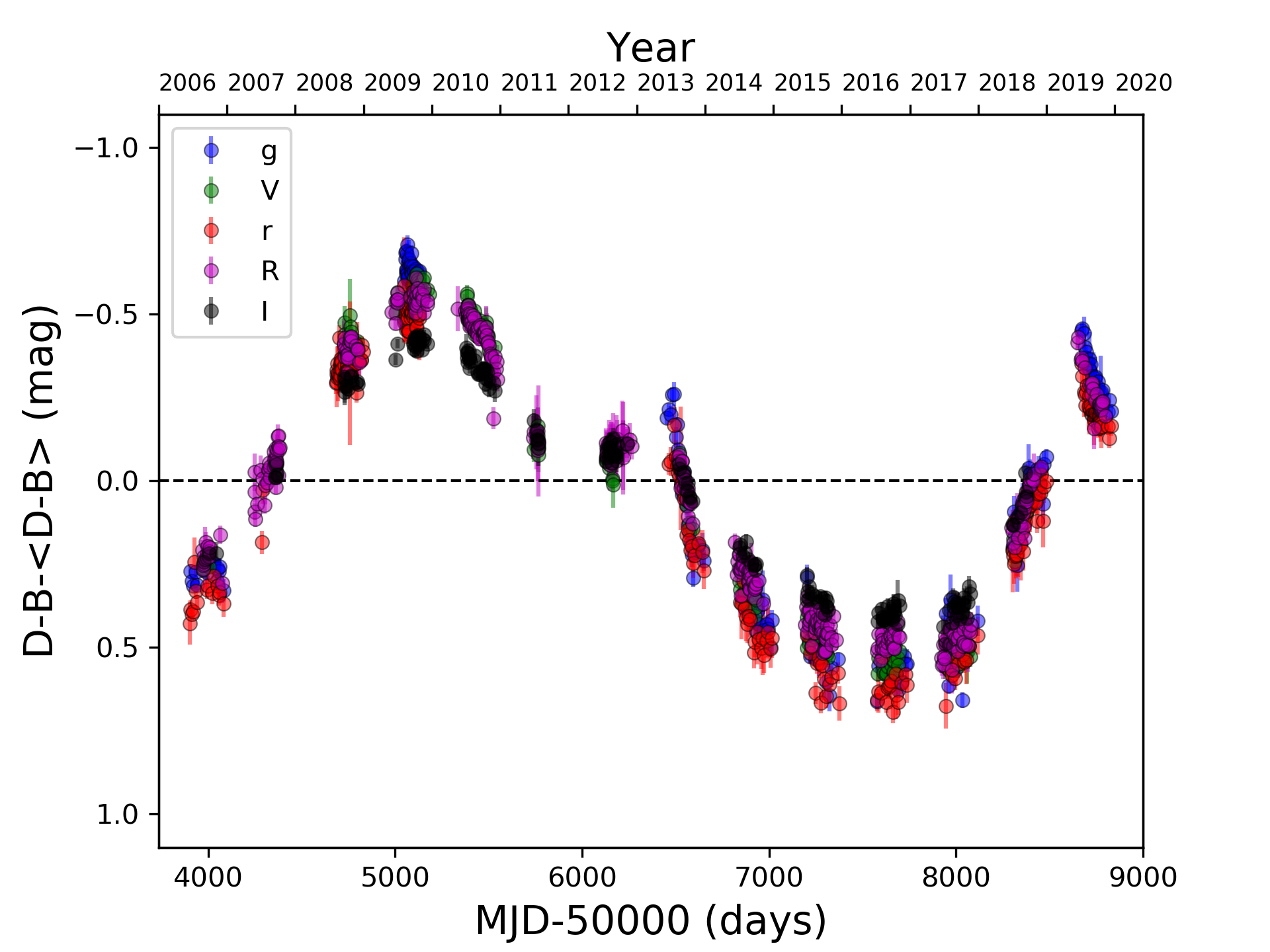}
\caption{Difference light curves of the Einstein Cross from the Liverpool-Maidanak 
brightness records in the $gVrRI$ bands. To construct these difference curves, the 
multi-band brightness records of B have been subtracted from those of A (top left 
panel), C (top right panel), and D (bottom panel). In addition to subtracting the 
records of 
the least variable image, the resulting difference curves are shifted in magnitude to show 
variations around the zero level.}
\label{fig:dcur}
\end{figure*}

As discussed in Sect.~\ref{sec:intro}, the chromatic microlensing 
variability of \object{QSO 2237+0305} can be analysed by obtaining magnitude 
differences between image 
pairs in several optical bands. Additionally, for this lens system, the source radius 
crossing and Einstein radius crossing timescales are 0.23 and 8.11 years, 
respectively \citep{Mosq11}. In view of the timescales involved, multi-band follow-up 
observations during 14 years might prove very useful for finding strong chromatic 
microlensing effects. To gain insight into the origin of variations in the quasar 
images that are most affected by microlensing, it is also convenient to use the light 
curves of 
the least variable image as reference records \citep[e.g.][]{Medi15}. Thus, we 
constructed difference light curves $A - B - \langle A - B \rangle$, $C - B - \langle 
C - B \rangle$, and $D - B - \langle D - B \rangle$ in the $gVrRI$ bands. These 
difference curves are depicted in Figure~\ref{fig:dcur}.

As expected, Figure~\ref{fig:dcur} shows significant microlensing variability in the
difference curves for A (top left panel), C (top right panel), and D (bottom panel).
Although the variability in $A - B$ and $D - B$ has a total amplitude of $\sim$ 1 mag, 
it is not as strong as in $C - B$. Chromaticity is also detected in these difference 
curves. For example, $D - B$ exhibits an oscillating global behaviour with an amplitude 
less than 1 mag in the $I$ band and more than 1 mag in the $g$ band. Moreover, $C - B$ 
include two sharp chromatic variations that occurred in the periods 2012$-$2013 and 
2015$-$2016. Although the whole set of difference curves can be used to probe the 
accretion disc structure and the composition of the lensing galaxy at relatively small 
impact parameters \citep[e.g.][]{Koch04,Eige08}, here we focus on the two prominent 
features of $C - B$. These resemble DCCE peaks seen in microlensing simulations, when 
sources enter regions interior to caustics (first crossing of a fold caustic) and 
later exit from them \citep[second caustic crossing; see e.g. Figs. 
10$-$11 of][]{Wamb98}. Based on this, each sharp variation of $C - B$ is thought to be 
due to 
a caustic crossing event in image C. The association between prominent variations in 
observed light curves and caustic crossings is sometimes justified by detailed 
numerical simulations \citep[e.g.][]{Gilm06,Angu08}.

\setcounter{table}{14}
\begin{table*}
\centering
\caption{Results from fitting caustic-crossing induced variations in 2012$-$2013.}
\begin{tabular}{lccccccccc}
\hline\hline
Band & \multicolumn{4}{c}{$p$ = 3/2 power law} & & 
\multicolumn{4}{c}{Gaussian} \\
\cline{2-5}
\cline{7-10}
 & $t_0$ (MJD$-$50\,000) & $\Delta t$ (d) & $c_{\rm{J}}/c_0$ & $\chi^2$/dof & & 
$t_0$ (MJD$-$50\,000) & $\Delta t$ (d) & $c_{\rm{J}}/c_0$ & $\chi^2$/dof \\
\hline 
$V$ & 6160.0 $\pm$ 1.2 & 52.6 $\pm$ 2.0 & 20.24 $\pm$ 0.33 & 1.69 & &
 6166.5 $\pm$ 1.5 & 90.1 $\pm$ 3.1 & 18.78 $\pm$ 0.30 & 1.68 \\ 
$R$ & 6160.2 $\pm$ 1.3 & 60.1 $\pm$ 2.1 & 18.37 $\pm$ 0.16 & 0.62 & &
 6164.7 $\pm$ 3.0 & 121.4 $\pm$ 5.9 & 17.51 $\pm$ 0.20 & 1.25 \\ 
$I$ & 6163.2 $\pm$ 2.0 & 79.3 $\pm$ 3.6 & 14.33 $\pm$ 0.16 & 1.20 & &
 6170.5 $\pm$ 4.3 & 146.2 $\pm$ 8.9 & 13.45 $\pm$ 0.25 & 2.49 \\ 
\hline
\end{tabular}
\tablefoot{
For each source profile ($p$ = 3/2 power law and Gaussian), the free parameters are 
the time of caustic crossing by the source centre ($t_0$), the source radius crossing 
time ($\Delta t$), and the relative caustic strength ($c_{\rm{J}}/c_0$). 
} 
\label{tab:cfratfit1}
\end{table*}

To demonstrate that the putative DCCE is a powerful tool for studying the accretion 
disc structure, we considered the corrected flux ratio 
$(C/B)_{\rm{corr}}$ over the 2009$-$2018 period for each optical band, that is, 
after removing a long-term 
microlensing gradient from $C/B$ (see Figure~\ref{fig:corfrat}). Our procedure is 
 discussed in depth and placed in perspective in Appendix~\ref{sec:appenb}. We analysed 
the two individual caustic-crossing events in a separate way. Unfortunately, we cannot 
draw the 2012$-$2013 event in the $g$ and $r$ bands because there is a long gap 
between days 5200 and 6400 in these passbands. Therefore, only the caustic-crossing 
induced variations in the $VRI$ bands were independently fitted to the three-parameter 
model $\mu_{\rm{caustic}} = 1 + (c_{\rm{J}}/c_0)J[(t - t_0)/\Delta t]$ for three 
different source profiles giving rise to analytical functions $J(z)$ (sources enter 
the caustic region; see Appendix~\ref{sec:appenb}). In general, the $p$ = 3/2 
power-law profile \citep{Shal01} leads to the best fits in terms of $\chi^2$/dof 
values, and the best-fit curves for $p$ = 3/2 power-law sources are shown in 
Figure~\ref{fig:corfrat} (solid lines). In Table~\ref{tab:cfratfit1} we also compare 
results for $p$ = 3/2 power-law sources and those for Gaussian sources. In each 
optical band, a $\Delta t$ (Gaussian) $\sim 2 \times \Delta t$ ($p$ = 3/2 power law) 
relationship is expected if both sources have the same half-light radius and move with 
the same velocity perpendicular to the caustic line \citep{Shal02}.    

\begin{figure}
\centering
\includegraphics[width=9cm]{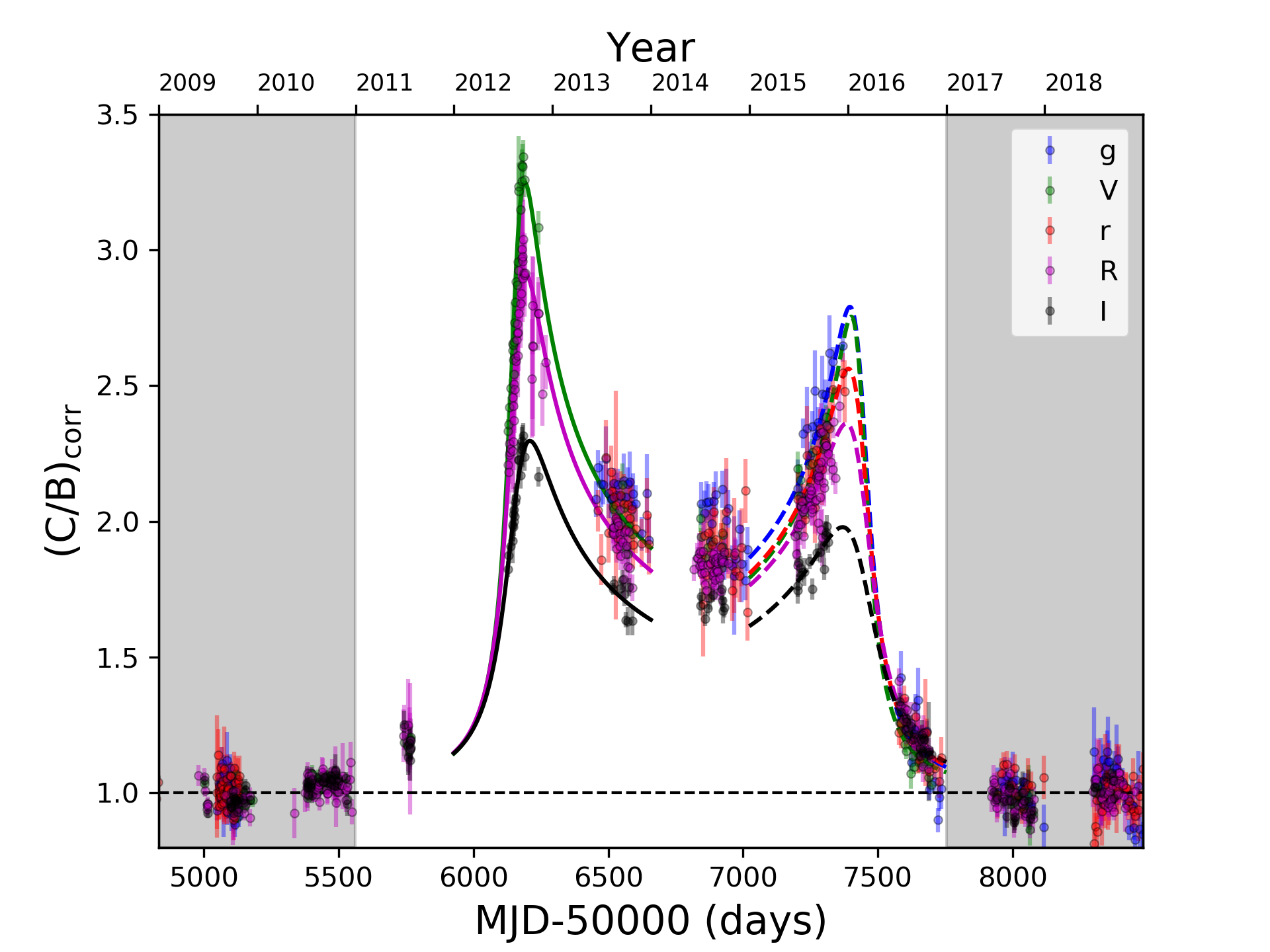}
\caption{Corrected flux ratio $(C/B)_{\rm{corr}}$ in the $gVrRI$ bands. Each sharp 
monochromatic variation is fitted to the microlensing model $\mu_{\rm{caustic}}$ for a
$p$ = 3/2 power-law source (see Appendix~\ref{sec:appenb}), yielding the solid and 
dashed lines (best-fit curves) associated with the caustic crossing events in 
2012$-$2013 and 2015$-$2016, respectively.}
\label{fig:corfrat}
\end{figure}

\begin{table}
\centering
\caption{Results from fitting caustic-crossing induced variations in 2015$-$2016 using
$p$ = 3/2 power-law sources.}
\begin{tabular}{lccc}
\hline\hline
Band & $\Delta t$ (d) & $c_{\rm{J}}/c_0$ & $\chi^2$/dof \\
\hline
$g$ &  63.3 $\pm$ 6.7 & 17.68 $\pm$ 0.42 & 3.18 \\  
$V$ &  55.0 $\pm$ 3.6 & 16.14 $\pm$ 0.22 & 1.60 \\
$r$ &  72.9 $\pm$ 8.2 & 16.55 $\pm$ 0.33 & 0.91 \\  
$R$ &  86.5 $\pm$ 4.4 & 15.71 $\pm$ 0.18 & 1.28 \\ 
$I$ & 110.4 $\pm$ 4.1 & 12.77 $\pm$ 0.14 & 0.84 \\ 
\hline
\end{tabular}
\tablefoot{
We obtain $t_0$ = 7433.1 $\pm$ 13.4 from data in the $R$ band, and only fit $\Delta t$ 
and $c_{\rm{J}}/c_0$ in each of the other four bands (see the notes to 
Table~\ref{tab:cfratfit1} for the meaning of the parameters). 
} 
\label{tab:cfratfit2}
\end{table}

Data in 2015$-$2016 make it possible to study the 
corresponding caustic-crossing event in all five bands. The microlensing model 
is now 
$\mu_{\rm{caustic}} = 1 + (c_{\rm{J}}/c_0)J[-(t - t_0)/\Delta t]$, which should 
provide $\Delta t >$ 0 values if sources actually exit from the caustic region. In 
order to improve results for a given source profile, we initially fit the best-sampled 
variation (in the $R$ band) to the model instead of performing five fits 
with three free parameters each. We then obtained $\Delta t$ and $c_{\rm{J}}/c_0$ in the
$gVrI$ bands, setting $t_0$ to that derived from $R$-band data. We reasonably assumed 
that all variations in $(C/B)_{\rm{corr}}$ are characterised by the same value of 
$t_0$. The lowest values of $\chi^2$/dof were again obtained by using the $p$ = 3/2 
power-law profile (see the dashed lines in Figure~\ref{fig:corfrat} and results in 
Table~\ref{tab:cfratfit2}). Consequently, our global results favour a surface 
brightness profile close to that of the standard accretion disc. As a general rule, 
profiles that most appreciably depart from the standard profile produce the worst fits, 
that is $p$ = 5/2 power law and Gaussian models. The global analysis also indicates that 
a DCCE most likely occurred. 

In addition, each of the two caustic crossings allows us to probe the relationship 
between source radius and emission wavelength. As a first indicator, we studied the 
correlation between $\Delta t$ and observed wavelength $\lambda_0$, where $\Delta t$ 
and $\lambda_0$ are proportional to $R_{\lambda}$ and $\lambda$, respectively (see 
Appendix~\ref{sec:appenb}). For the first caustic crossing event in 2012$-$2013, we 
considered solutions of $\Delta t$ for $p$ = 3/2 power law and Gaussian source models. 
In addition to the results through $(C/B)_{\rm{corr}}$ (see Table~\ref{tab:cfratfit1}), 
new solutions were inferred from $(C/A)_{\rm{corr}}$. This complementary analysis is 
useful to determine the influence of the choice of smoothly varying reference records on 
the $\Delta t-\lambda_0$ relationship. For the second caustic crossing event in 
2015$-$2016, we only used the results via $(C/B)_{\rm{corr}}$ in 
Table~\ref{tab:cfratfit2} ($p$ = 3/2 power-law sources).
 
\begin{figure}
\centering
\includegraphics[width=9cm]{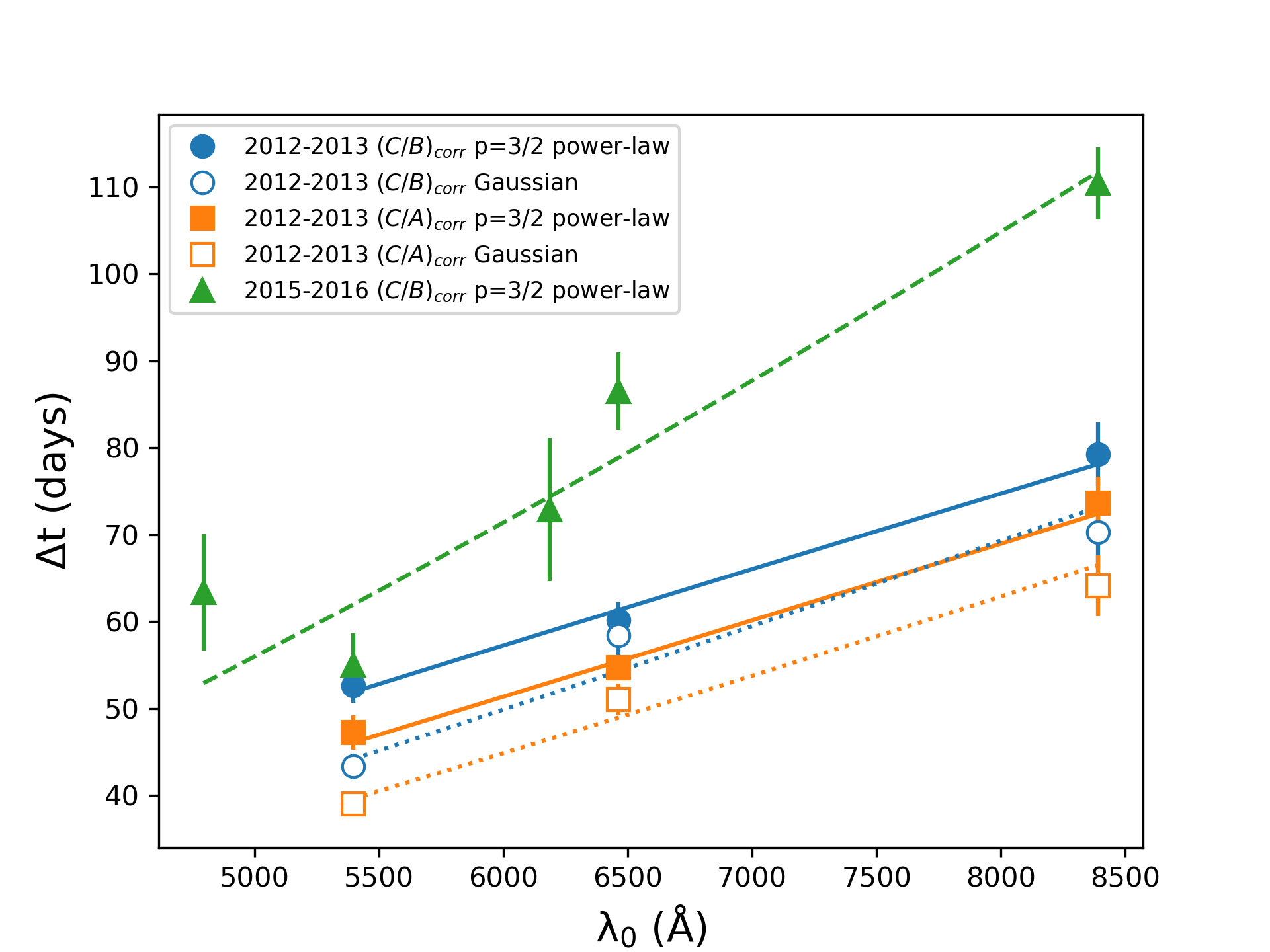}
\caption{Chromatic behaviour of the source radius crossing time during the 2012$-$2013 
and 2015$-$2016 microlensing events in image C. The $\Delta t$ values for Gaussian 
sources are reduced by a factor of about 2 for comparison purposes, and the five lines
decribe power-law fits (see main text).}
\label{fig:radwave}
\end{figure}

Figure~\ref{fig:radwave} displays the chromatic behaviour of $\Delta t$, along with 
power-law fits. These fits inform us about the power-law index $\alpha$ in 
$R_{\lambda} \propto \lambda^{\alpha}$ (see Table~\ref{tab:alpha}). The 2015$-$2016 
event leads to radii of sources at five emission wavelengths that are consistent with 
a standard accretion disc ($\alpha$ = 4/3), although both $\chi^2$/dof and the scatter 
in $\alpha$ values are relatively high. This measure ($\alpha$ = 1.34 $\pm$ 0.23) 
confirms but does not improve a previous result through ESO-OGLE multi-band light 
curves of \object{QSO 2237+0305} \citep[$\alpha$ = 1.2 $\pm$ 0.3;][]{Eige08}. 
When we consider our best power-law fits to radii at three wavelengths for the 
event in 2012-2013 ($\chi^2$/dof $\leq$ 1), the index is measured to $\sim$10\% 
precision: 
$\alpha$ = 1.0 $\pm$ 0.1. The new measurement is more accurate than previous estimates 
based on chromatic variations of the Einstein Cross \citep[e.g.][]{Eige08,Mun16},
suggesting that as the emission wavelength increases, the source radius grows more 
smoothly than the standard disc radius.  

\begin{table}
\centering
\caption{Power-law index of the radius-wavelength relation $R_{\lambda} \propto 
\lambda^{\alpha}$ via the chromaticity of $\Delta t$.}
\begin{tabular}{lccc}
\hline\hline
Method & Source profile & $\alpha$ & $\chi^2$/dof \\
\hline
\multicolumn{4}{c}{2012$-$2013 event} \\
\hline
$(C/B)_{\rm{corr}}$ & $p$ = 3/2 power law & 0.93 $\pm$ 0.10 & 0.57 \\  
                    & Gaussian            & 1.14 $\pm$ 0.26 & 2.86 \\   
$(C/A)_{\rm{corr}}$ & $p$ = 3/2 power law & 1.02 $\pm$ 0.12 & 0.87 \\                         
                    & Gaussian            & 1.17 $\pm$ 0.19 & 2.15 \\ 
\hline
\multicolumn{4}{c}{2015$-$2016 event} \\
\hline    
$(C/B)_{\rm{corr}}$ & $p$ = 3/2 power law & 1.34 $\pm$ 0.23 & 3.07 \\                               
\hline
\end{tabular}
\label{tab:alpha}
\end{table}

\begin{figure}
\centering
\includegraphics[width=9cm]{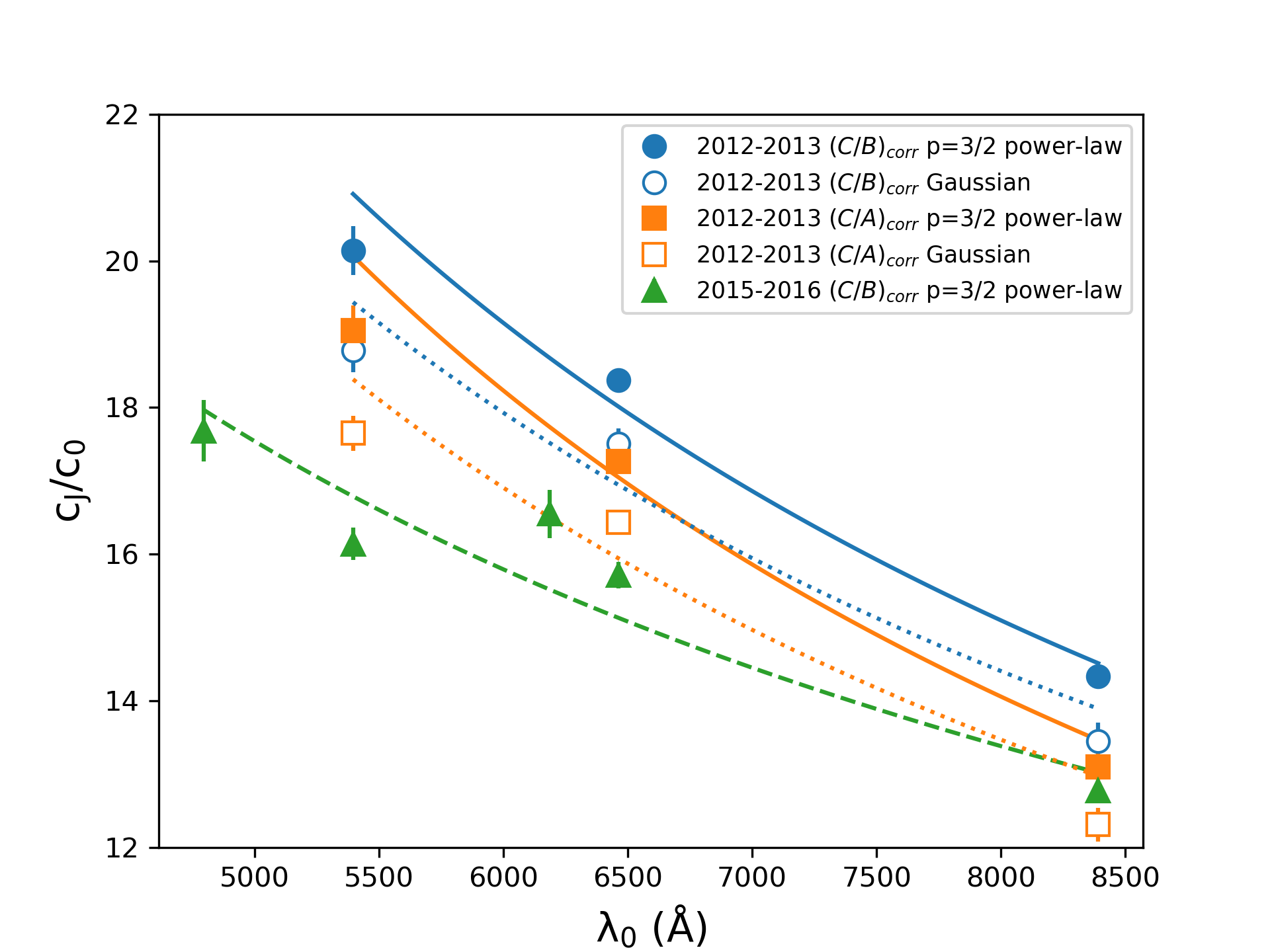}
\caption{Chromatic behaviour of the relative caustic strength during the 2012$-$2013 
and 2015$-$2016 microlensing events in image C. The five lines represent power-law 
fits (see main text).}
\label{fig:radwave2}
\end{figure}

\begin{table}
\centering
\caption{Power-law index of the radius-wavelength relation $R_{\lambda} \propto 
\lambda^{\alpha}$ via the chromaticity of $c_{\rm{J}}/c_0$.}
\begin{tabular}{lcc}
\hline\hline
Method & Source profile & $\alpha$ \\
\hline
\multicolumn{3}{c}{2012$-$2013 event} \\
\hline
$(C/B)_{\rm{corr}}$ & $p$ = 3/2 power law & 1.66 $\pm$ 0.28 \\  
                    & Gaussian            & 1.52 $\pm$ 0.43 \\   
$(C/A)_{\rm{corr}}$ & $p$ = 3/2 power law & 1.81 $\pm$ 0.39 \\                         
                    & Gaussian            & 1.58 $\pm$ 0.54 \\ 
\hline
\multicolumn{3}{c}{2015$-$2016 event} \\
\hline    
$(C/B)_{\rm{corr}}$ & $p$ = 3/2 power law & 1.15 $\pm$ 0.22 \\                               
\hline
\end{tabular}
\label{tab:alpha2}
\end{table}

In principle, we can use a second $\alpha$ indicator. The relative caustic strength is 
proportional to 1/$R_{\lambda}^{1/2}$, and the chromaticity of $c_{\rm{J}}/c_0$ 
therefore provides complementary information on $\alpha$. As shown in 
Figure~\ref{fig:radwave2},
the residuals from power-law fits are very large, yielding uncomfortable $\chi^2$/dof 
values $\geq$ 10. Therefore we did not include these values in Table~\ref{tab:alpha2}
and do not try to discuss results individually. However, the set of measures of 
$c_{\rm{J}}/c_0$ indicates that the relative caustic strength really decreases as the 
radius (wavelength) increases, and we considered all estimates of $\alpha$ in 
Tables~\ref{tab:alpha} and \ref{tab:alpha2}, ignoring differences in errors and 
$\chi^2$/dof values, to compute the mean and its standard deviation: 1.33 $\pm$ 0.09.
This statistical approach clearly favours a standard accretion disc.
        
\section{Conclusions}
\label{sec:end}

Within the framework of a collaboration between the GLENDAMA team and several groups 
operating the Maidanak Observatory, we analysed a joint database of the 
quadruply imaged quasar \object{QSO 2237+0305}. This large database contains optical 
frames in the $gVrRI$ bands in the period 2006$-$2019, which were used to extract 
fluxes of the four quasar images A-D in a homogeneous way. Our 14-year multi-band 
light curves are expected to appreciably contribute to a better knowledge of the 
structure of the accretion disc around the central supermassive black hole, as well as 
the composition of the lensing galaxy bulge \citep[e.g.][]{Koch04,Eige08}. We 
concentrated on two sharp chromatic microlensing events that appear in the 
difference light curve $C - B$. These prominent features in 2012$-$2013 and 2015$-$2016 
were tentatively associated with two consecutive caustic-crossing events in image C.

We studied the two putative caustic-crossing events in a separate way, considering 
three different models for the brightness profile of involved sources: the widely
used Gaussian model, and the $p$ = 5/2 and $p$ = 3/2 power-law models \citep{Shal02}.
This choice allowed us to probe the behaviour of source profiles close to that of the 
standard accretion disc ($p$ = 3/2), profiles that noticeably depart from it 
(Gaussian), and intermediate profiles ($p$ = 5/2). Our global results favour the 
$p$ = 3/2 power-law profile. The global analysis also supports our initial hypothesis 
of a double caustic-crossing event in image C. While previous studies reported
isolated strong-microlensing episodes in images of the Einstein Cross \citep[e.g.][and 
references therein]{Medi15}, the Liverpool-Maidanak light curves suggest that image C 
has been affected by a double caustic crossing between 2012 and 2016. More 
specifically, our results are consistent with a standard accretion disc entering a region 
interior to a caustic and then exiting from it. However, we note that despite these 
encouraging conclusions, a final confirmation of the DCCE requires a description of 
the full microlensing event (including its central part) through numerical simulations 
\citep[e.g.][]{Wamb98,Koch04}. Such detailed modelling is beyond the scope of this 
paper (see below).

We also probed the relationship between source radius and emission wavelength: 
$R_{\lambda} \propto \lambda^{\alpha}$ at $\lambda \sim 1780-3110$ \AA, using two 
$\alpha$ indicators. For these sources of UV radiation, $\alpha$ is unfortunately not 
as well constrained as would be desirable. The statistical 
result based on ten solutions from both indicators contradicts the measurement based 
on the two 
best solutions with $\chi^2$/dof $\leq$ 1. This last measurement of $\alpha$ (1.0 
$\pm$ 0.1) relies on only one indicator and $VRI$ data during the 2012$-$2013 event. 
On the other hand, the statistical result from both indicators ($\alpha$ = 1.33 $\pm$ 
0.09) takes into account $VRI$ data during the first event and $gVrRI$ data over the 
second event, although most individual solutions have $\chi^2$/dof $>$ 1. This 
estimate is fully consistent with a standard accretion disc ($\alpha$ = 4/3), and it 
is probably more representative (unbiased) than the other. Again we need to perform 
numerical microlensing simulations to robustly constrain the value of $\alpha$, which 
has been measured from microlensing-induced chromatic variations to $\sim$25$-$30\% 
precision \citep{Eige08,Mun16}, or to $\sim$10\% formal precision, but in an 
ambiguous way (this paper).

A deep analysis of the Liverpool-Maidanak light curves with the aid of detailed 
microlensing simulations will be presented in a subsequent paper. The future paper 
will focus on, among other things, providing a probabilistic confirmation of the DCCE 
in image C, discussing the brightness profile of UV continuum sources, and 
constraining the size and structure of the accretion disc. When we were completing 
this paper, the COSmological MOnitoring of GRAvItational Lenses (COSMOGRAIL) 
collaboration reported ESO $R$-band light curves of \object{QSO 2237+0305} 
that cover the period 2010$-$2013 \citep[see Fig. B.11 of][]{Mill20}. Although these 
COSMOGRAIL light curves are not yet publicly available and were obtained using a 
photometric approach different from ours, they might play a role in the future 
analysis through numerical microlensing simulations.      

\begin{acknowledgements}
This paper is based on observations made with the Liverpool Telescope (LT) and the 
AZT-22 Telescope at the Maidanak Observatory (MT). The LT is operated on the island of 
La Palma by Liverpool John Moores University in the Spanish Observatorio del Roque de 
los Muchachos of the Instituto de Astrofisica de Canarias with financial support from 
the UK Science and Technology Facilities Council. We thank the staff of the LT for a 
kind interaction before, during and after the observations. The Maidanak Observatory 
is a facility of the Ulugh Beg Astronomical Institute (UBAI) of the Uzbekistan Academy 
of Sciences, which is operated in the framework of scientific agreements between UBAI 
and Russian, Ukrainian, US, German, French, Italian, Japanese, Korean, Taiwan, Swiss 
and other countries astronomical institutions. We thank O. Ye. Kochetov and V. V. 
Konichek for performing some observations with the MT. We also used data taken from 
the Sloan Digital Sky Survey (SDSS) web site (\url{http://www.sdss.org/}), and we are 
grateful to the SDSS collaboration for doing that public database. This research has 
been supported by the MINECO/AEI/FEDER-UE grant AYA2017-89815-P, the University of 
Cantabria, and the Ministry of Innovative Development of Uzbekistan grants F2-FA-F026 
and VA-FA-F-2-010.
\end{acknowledgements}

\clearpage

\begin{appendix}

\section{Magnitudes of quasar images using different photometric techniques and 
telescopes}
\label{sec:appena}

Here, we compare our MT $V$-band data at 19 epochs in 2006$-$2008 with MT $V$-band 
magnitudes at the same epochs derived from a different photometric approach 
\citep[][henceforth D10]{Dudi10}, and also with concurrent OGLE data\footnote{OGLE 
photometry is available at \url{http://ogle.astrouw.edu.pl/}}. The OGLE collaboration 
monitored \object{QSO 2237+0305} in the $V$ band from 1997 to the first half of 2009, 
using a photometric technique different from ours and a telescope in the southern 
hemisphere \citep{Wozn00,Udal06}. Briefly, we used the software 
IMFITFITS \citep{McLe98} to perform PSF-fitting photometry, modelling the 
lensing galaxy bulge as a de Vaucouleurs profile convolved with the PSF and taking 
HST astrometric constraints into account (see Sect.~\ref{sec:obslcur}). \citet{Vaku04} 
introduced another flux-extraction technique that was also applied to MT frames in the 
period of interest \citepalias{Dudi10}. In this alternative PSF photometry, for 
instance, the lensing galaxy is described as the sum of three elliptical Gaussian 
functions. Additionally, the OGLE light curves rely on the so-called image subtraction 
method \citep[e.g.][]{Alar98}. After the OGLE team obtained differences between 
individual frames 
and a reference stacked image, differential quasar fluxes were measured through 
PSF-fitting photometry with HST astrometric constraints \citep[e.g.][]{Udal06}.   

\begin{figure}
\centering
\includegraphics[width=9cm]{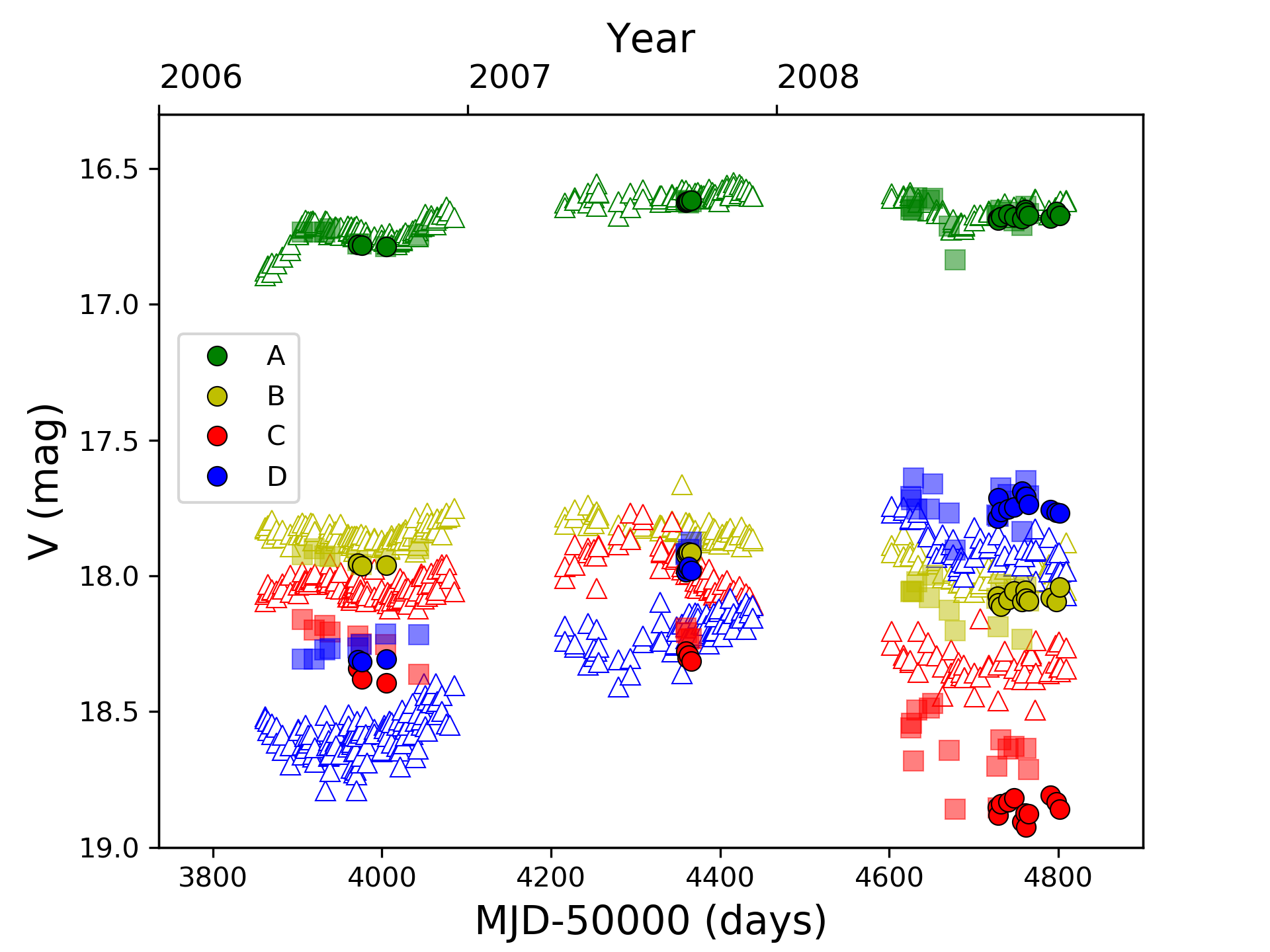}
\caption{Comparison of $V$-band magnitudes in 2006$-$2008 using different photometric 
approaches and telescopes. Our MT data (circles), MT data from an alternative flux 
extraction technique (squares), and OGLE brightness records (triangles).}
\label{fig:compar}
\end{figure}

In Figure~\ref{fig:compar} we display the measures in this paper (circles; see Table 
11), MT data from a flux extraction technique different to ours 
\citepalias[squares;][]{Dudi10}, and OGLE light curves (triangles). MT magnitudes from 
both photometric approaches are remarkably similar for the brightest images (see 
Table~\ref{tab:moffset}). However, there are significant magnitude offsets with 
respect to OGLE data, in particular for the faintest images. We note that the 
comparison in this appendix can be particularly useful for studying 
25-year $V$-band records of the four quasar images because MT and OGLE observations 
cover the period 1995$-$2019. 

\begin{table}
\centering
\caption{Mean magnitude offsets.}
\begin{tabular}{lcccc}
\hline\hline
 & A & B & C & D \\
\hline
This paper$-$D10  & 0.004 & $-$0.004 & 0.098 &    0.036 \\
This paper$-$OGLE & 0.037 &    0.104 & 0.422 & $-$0.240 \\
\hline
\end{tabular}
\tablefoot{
Mean magnitude offsets are estimated based on a certain number of concurrent observing 
epochs, i.e. 15 out of 19 for the comparison with \citetalias{Dudi10} and 10 out of 19 
for the comparison with OGLE. 
} 
\label{tab:moffset}
\end{table}

\section{Analysis of caustic crossing events in image C of QSO 2237+0305}
\label{sec:appenb}

When the continuum source emitting at $\lambda$ crosses a fold caustic for a given 
quasar image, the observed flux of such an image at $\lambda_0$ = $\lambda (1 + 
z_{\rm{s}})$ changes 
dramatically over time. Because the accretion disc of \object{QSO 2237+0305} is seen 
face-on \citep[e.g.][]{Poin10}, it is reasonable to consider an axisymmetric source to 
model this flux variation during the caustic-crossing event. Moreover, the fold 
caustic is generally assumed to be a straight line, so that the theoretical microlensing 
curve can be primarily built by convolving the axisymmetric intensity profile with the 
straight-fold magnification \citep[e.g.][]{Schn87,Shal02}. However, there are a number 
of caveats concerning this simple approximation, which does not work properly in some 
cases. For example, \citet{Fluk99} noted that the model is only correct when the source 
size is much smaller than that of the fold caustic, while \citet{Gaud02} incorporated 
a slowly (linearly) varying background term \citep[see also][]{Yone01}. Very recently, 
\citet{Weis19} have also discussed the magnification of a source near a fold caustic 
using higher order approximations and numerical simulations. Interestingly, higher 
order approximations were required to accurately fit a caustic-crossing flux variation 
($V$ band) in image C of the Einstein Cross that occurred in 1999 \citep{Alex11}. For 
a Gaussian brightness profile, the $\chi^2$/dof value decreased from 1.35 to 0.9$-$1 
when suitable corrections to the straight-fold model were taken into account. 

Whereas \citet{Alex11} have neglected possible intrinsic variations in image C, 
\citet{Medi15} have removed intrinsic fluctuations of images A and C by dividing their 
$V$-band caustic-crossing flux variations by flux records of the least variable 
images. \citet{Medi15} also showed that the global shape of these flux ratios  
(associated with three caustic-crossing events occurring before 2006 or ending in 
2006) can be reasonably fitted by the straight-fold model plus a linearly varying 
background contribution. In more detail, when the intensity profile of the standard 
thin disc is used \citep{Shak73}, the $\chi^2$/dof values are $\sim$1.5$-$3.6. 
Although these 
fits can be easily improved by considering a more realistic magnification pattern for 
the image undergoing the caustic crossing \citep{Alex11}, a microlensing magnification 
gradient for the image with the smoothest variability and/or a different source 
profile \citep[e.g. including relativistic effects;][]{Abol12,Medi15}, they are enough 
to prove the great potential of the Einstein Cross light curves.  

We focused on two new caustic-crossing events in image C of \object{QSO 
2237+0305}, which occurred in 2012$-$2013 and 2015$-$2016 (see 
Sect.~\ref{sec:diflcur}). When the difference curves $C - B$ are used to build the flux 
ratio $C/B$ in the $gVrRI$ bands over the 2009$-$2018 period, it is apparent that this 
ratio contains a long-term microlensing gradient (see the top right panel of 
Fig.~\ref{fig:dcur}). This gradient originates in the image that experiences 
caustic crossings \citep[C; e.g.][]{Gaud02} or in the least variable image B. In the 
first scenario, each caustic-crossing induced variation in $C/B$ was modelled as
 \begin{equation}
    f_{\lambda} = \frac{c_0 + c_1(t - t_1) + c_{\rm{J}}J[(t - t_0)/T]}{b_0} \,,
 \label{eq1}
 \end{equation}
where $c_0 + c_{\rm{J}}J[(t - t_0)/T]$ results from the convolution of the source 
profile with the straight-fold magnification \citep[straight-fold model; see, e.g., 
Fig. 1 and Eq. (13) in][]{Shal02}. In Eq.~\ref{eq1}, $c_0$ and $c_1$ are the constant 
background and slope of the linear gradient for image C ($t_1$ is an epoch in the 
2009$-$2018 period that can be conveniently fixed), $c_{\rm{J}}$ is related to the 
caustic strength, $t_0$ is the time of caustic crossing by the source centre, $T = \pm 
\Delta t$ (the plus indicates that the source enters the caustic region, while the minus 
denotes that the source exits from it; $\Delta t = R_{\lambda}/V_{\perp}$ 
is the source radius crossing time, and $R_{\lambda}$ and $V_{\perp}$ are the typical 
radius of the source profile and source velocity perpendicular to the caustic line), 
and $b_0$ is the constant term for image B (we implicitly assume that B does not 
suffer a time-varying microlensing magnification).   

For the second scenario, instead of Eq.~\ref{eq1}, we used
 \begin{equation}
    f_{\lambda} = \frac{c_0 + c_{\rm{J}}J[(t - t_0)/T]}{b_0 + b_1(t - t_1)} \,,
 \label{eq2}
 \end{equation}
where $b_1$ is the slope of the linear gradient for B. Equations~\ref{eq1} 
and Eq.~\ref{eq2} show that both models have the same number of parameters. 
We considered three axisymmetric shapes for the source profiles that produce 
analytical functions 
$J(z)$. These are the Gaussian model, and the $p$ = 5/2 and $p$ = 3/2 power-law models
introduced by \citet{Shal01}. As shown in Fig. 4 of \citet{Shal02}, the $p$ = 3/2 
power-law profile closely mimics the behaviour of the standard accretion disc, and the 
Gaussian profile departs more strongly from the standard behaviour. After some
initial tests, we realised that the second secenario, that is, Eq.~\ref{eq2}, fits the 
caustic-crossing induced variations in $C/B$ better. We therefore chose to 
present results for this theoretical microlensing model. The reference epoch $t_1$ was
set to 1 January 2014, and the $C/B$ values in the 2009$-$2010 and 2017$-$2018 periods 
(when sources are presumably not affected, or very weakly affected, by the caustic 
region and the $J$ function becomes zero or negligible) were used to determine 
$c_0/b_0$ and $b_1/b_0$ (see Table~\ref{tab:micgrad}). After these parameters
that are not related to caustic effects are derived, the key idea is to remove the 
long-term microlensing 
gradient by computing the corrected flux ratio $(C/B)_{\rm{corr}} = (C/B) \{ [1 + 
(b_1/b_0)(t - t_1)]/(c_0/b_0) \}$ in the $gVrRI$ bands, and then fit each 
caustic-crossing induced variation in $(C/B)_{\rm{corr}}$ to the microlensing law 
$\mu_{\rm{caustic}} = 1 + (c_{\rm{J}}/c_0)J[(t - t_0)/T]$ (see 
Sect.~\ref{sec:diflcur}). In addition to $\Delta t \propto R_{\lambda}$, it is 
straightforward to show that $c_{\rm{J}}/c_0 \propto 1/R_{\lambda}^{1/2}$.     

\begin{table}
\centering
\caption{Long-term microlensing gradient in $C/B$.}
\begin{tabular}{cccccc}
\hline\hline
Parameter & $g$ & $V$ & $r$ & $R$ & $I$ \\
\hline
$c_0/b_0$ & 0.355 & 0.371 & 0.357 & 0.375 & 0.431 \\
$b_1/b_0$ & 0.320 & 0.311 & 0.340 & 0.344 & 0.294 \\
\hline
\end{tabular}
\tablefoot{
Using 2009$-$2010 and 2017$-$2018 data, we interpret the long-term microlensing gradient in $C/B$ 
 as due to linearly varying microlensing in image B. Here, $c_0/b_0$ is 
a dimensionless parameter and $b_1/b_0$ is a relative slope in 10$^{-3}$ day$^{-1}$ 
(see main text).  
} 
\label{tab:micgrad}
\end{table}
         
\end{appendix}

\end{document}